\documentclass[]{elsarticle}

\usepackage{amsmath,amssymb,color,graphics,graphicx,psfrag,url,floatrow,afterpage,wrapfig}

\usepackage{natbib}
\usepackage{float}
\usepackage{subfig}
\usepackage{color}
\usepackage{multirow, lscape, bm}

\newcommand   \D [2]{\frac{\partial   #1}{\partial #2}}

\renewcommand{\vec}[1]{\bm{\mathrm{#1}}}

\def \CC{\mathbb{C}}

\def \FF{\mathbb{F}}
\def \II{\mathbb{I}}
\def \PPe{\mathbb{P}^{\text{s}}}

\def \N{\vec{N}}

\def \X{\vec{X}}
\def \f{\vec{f}}

\def \n{\vec{n}}
\def \s{\vec{s}}

\def \u{\vec{u}}

\def \vchi{\vec{\chi}}
\def \x{\vec{x}}
\def \grad{\nabla}

\def \p{\partial}

\def \dA{\mbox{d} A(\X)}
\def \dX{\mbox{d} \X}

\def \dx{\mbox{d} \x}

\def \sigmae{\vec{\sigma}^{\text{s}}}

\def \sigmaf{\vec{\sigma}^{\text{f}}}
\def \vsigma{\vec{\sigma}}

\def \fe{\f^{\text{s}}}

\def \Ca{\text{Ca}}

\journal{Journal of Medical Engineering \& Physics}

\begin{document}

\begin{frontmatter}

\title{A coupled mitral valve -- left ventricle model with fluid-structure interaction}

\author{Hao Gao}
\address{School of Mathematics and Statistics, University of Glasgow, UK}

\author{Liuyang Feng}
\address{School of Mathematics and Statistics, University of Glasgow, UK}

\author{Nan Qi}
\address{School of Mathematics and Statistics, University of Glasgow, UK}

\author{Colin Berry}
\address{Institute of Cardiovascular and Medical Science, University of Glasgow, UK}
\author {Boyce Griffith}
\address{Departments of Mathematics and Biomedical Engineering and McAllister Heart Institute, University of North Carolina, Chapel Hill, NC, USA}
\author{Xiaoyu Luo}
\address{School of Mathematics and Statistics, University of Glasgow, UK}

\begin{abstract}
Understanding the interaction between the valves and walls of the heart is important in assessing and subsequently treating heart dysfunction.  With advancements in cardiac imaging, nonlinear mechanics and computational techniques, it is now possible to explore the mechanics of valve-heart interactions using anatomically and physiologically realistic models. This study presents an integrated model of the mitral valve (MV) coupled to the left ventricle (LV), with the geometry derived from in vivo clinical magnetic resonance images. Numerical simulations using this coupled MV-LV model are developed using an immersed boundary/finite element method.  The model incorporates detailed valvular features, left ventricular contraction, nonlinear soft tissue mechanics, and fluid-mediated interactions between the MV and LV wall.  We use the model to simulate the cardiac function from diastole to systole, and investigate how myocardial active relaxation function affects the LV pump function.   The results of the new model agree with in vivo measurements, and demonstrate that the diastolic filling pressure increases significantly with impaired myocardial active relaxation to maintain the normal cardiac output.   The coupled model has the potential to advance fundamental knowledge of mechanisms underlying MV-LV interaction, and help in risk stratification and optimization of therapies for heart diseases. 
\end{abstract}


\end{frontmatter}
\clearpage

\section{Introduction}
The mitral valve (MV) has a complex structure that includes two distinct asymmetric leaflets, a mitral annulus, and chordal tendinae that connect the leaflets to papillary muscles that attach to the wall of the left ventricle (LV).  MV dysfunction remains a major medical problem because of its close link to cardiac dysfunctions leading to morbidity and premature mortality~\citep{go2014heart}.  

Computational modelling for understanding the MV mechanics promises more effective MV repairs and replacement~\cite{sacks2009biomechanics,votta2013toward,sun2014computational,kheradvar2015emergingSolutionForFuture}.  Biomechanical MV models have been developed for several decades, starting from the simplified two-dimensional approximation to three-dimensional models, and to multi-physics/-scale models~\cite{kunzelman1992stress,dahl2012fsi,weinberg2010multiscale,wang2013finite,Prot-Skallerud2009,Stevanella-M2011,lee2015quantification}.  Most of previous studies were based on structural and quasi-static analysis applicable to a closed valve~\cite{einstein2010fluid}; however, MV function during the cardiac cycle  cannot be fully assessed without modelling the ventricular dynamics and the fluid-structure interaction (FSI) between the MV, ventricles, and the blood flow~\cite{einstein2010fluid,gao2016modelling}. 


Because of the complex interactions among the MV, the sub-mitral apparatus, the heart walls, and the associated blood flow,  few modelling studies have been carried out that  integrate the MV and ventricles in a single model~\cite{wenk2010first,wong2012effect,baillargeon2015human}.   Kunzelman, Einstein, and co-workers first simulated normal and pathological mitral function~\cite{Einstein2003, Einstein2005,kunzelman2007fluid} with FSI using LS-DYNA (Livermore Software Technology Corporation, Livermore, CA, USA) by putting the MV into a straight tube. Using similar modelling approach, Lau et al.~\citep{lau2010mitral} compared MV dynamics with and without FSI, and they found that valvular closure configuration is different when using the FSI MV model. Similar findings are reported by Toma et al~\cite{toma2016fluidFSI}.  Over the last few years, there have also been a number of FSI valvular models using the immersed boundary (IB) method to study the flow across the MV~\cite{Watton2008,ma2012image,gao2014finite}. In a series of studies, Toma~\cite{toma2015fluid,toma2016fluidFSI,toma2016fluid} developed a FSI MV model based on in vitro  MV experimental system to study the function of the chordal structure, and good agreement was found between the computational model and in vitro experimental measurements. However, none of the aforementioned MV models accounted for the MV interaction with the LV dynamics. Indeed, Lau et al.~\cite{lau2010mitral} found that even with a fixed U-shaped ventricle, the flow pattern is substantially different from that estimated using a tubular geometry. Despite the advancements in computational modelling of individual MV~\cite{einstein2010fluid,lee2015quantification} and LV models~\cite{nash2000computational,Chen2016,quarteroni2016integrated}, it remains challenging to develop an integrated MV-LV model which includes the strong coupling between the valvular deformation and the blood flow. Reasons for this include  limited data for model construction, difficult choices of boundary conditions, and large computational resources required by these simulations. 

Wenk et al. \cite{wenk2010first} reported a structure-only MV-LV model using LS-DYNA that included the LV, MV, and chordae tendineae. This model was later extended to study MV stress distributions using a saddle shaped and asymmetric mitral annuloplasty ring~\cite{wong2012effect}.  A more complete whole-heart model was recently developed using a human cardiac function simulator in the Dassault Systemes's \emph{Living Heart} project \cite{baillargeon2015human}, which includes four ventricular chambers, cardiac valves, electrophysiology, and detailed myofibre and collagen architecture. Using the same simulator, effects of different mitral annulus ring were studied by Rausch et al.~\cite{rausch2016virtual}. However, this simulator does not yet account for detailed FSI. 

The earliest valve-heart coupling model that includes FSI is credited to Peskin and McQueen's pioneering work in the 1970s \cite{peskin1972flow, mcqueen1982fluid,peskin1977numerical} using the classical IB approach~\cite{Peskin2002}.   Using this same method, Yin et al. \cite{Yin2010} investigated fluid vortices associated with the LV motion as a prescribed moving boundary.  Recently, Chandran and Kim~\cite{chandran2015computational} reported a prototype FSI MV dynamics in a simplified LV chamber model during diastolic filling using an immersed interface-like approach.  One of the key limitations of these coupled models is the simplified representation of the biomechanics of the LV wall.  To date, there has been no work reported a coupled MV-LV model which has full FSI and based on realistic geometry and experimentally-based models of soft tissue mechanics. 
 
This study reports an integrated MV-LV model with FSI derived from in vivo images of a healthy volunteer.  Although some simplifications are made, this is the first three-dimensional FSI MV-LV model that includes MV dynamics, LV contraction, and experimentally constrained descriptions of nonlinear soft tissue mechanics. This work is built on our previous models of the MV  \cite{ma2012image,gao2014finite} and LV~\cite{gao2014quasi,Chen2016}.   The model is implemented using a hybrid immersed boundary method with finite element elasticity (IB/FE) \cite{Griffith2012-FEIB}. 

\section{Methodology}
\subsection{IB/FE Framework}
The coupled MV-LV model employs an Eulerian description for the blood, which is modelled as a viscous incompressible fluid, along with a Lagrangian description for the structure immersed in the fluid. The fixed physical coordinates are $\x = (x_1,x_2,x_3) \in \Omega$, and the Lagrangian reference coordinate system is $\X = (X_1,X_2,X_3) \in U$. The exterior unit normal along $\partial U$ is $\N(\X)$. Let $\vchi(\X,t)$ denote the physical position of any material point $\X$ at time $t$, so that ${\vchi(U,t)}=\Omega^\text{s}(t)$ is the physical region occupied by the immersed structure.  The IB/FE formulation of the FSI system reads 
\begin{align}
\rho\left(\D{\u}{t}(\x,t) + \u(\x,t) \cdot \grad \u(\x,t)\right)
&= - \grad p(\x,t) + \mu \grad^2 \u(\x,t) + \fe(\x,t),       \label{eqn::ns}                                 \\
\grad \cdot \u(\x,t) &= 0,      \label{eqn::imcompressibility}  
\\
\fe(\x,t)            &= \int_U \grad \cdot \PPe(\X,t) \, \delta(\x - \vec{\chi}(\X,t)) \, \dX                 \nonumber                   \\
& \ \ \mbox{} - \int_{\p U} \PPe(\X,t) \, \N(\X) \, \delta(\x - \vec{\chi}(\X,t)) \, \dA,   \label{eqn::interact1}   \\
\D{\vec{\chi}}{t}(\X,t) &= \int_\Omega \u(\x,t) \, \delta(\x - \vec{\chi}(\X,t)) \, \dx,     \label{eqn::interact2}                       
\end{align}
where $\rho$ is the fluid density, $\mu$ is the fluid viscosity, $\u$ is the Eulerian velocity, $p$ is the Eulerian pressure, and $\fe$ is the Eulerian elastic force density. Different from the classical IB approach \cite{Peskin2002}, here the elastic force density $\fe$ is determined from the first Piola-Kirchoff stress tensor of the immersed structure $\PPe$ as in Eq.~\ref{eqn::interact1}.  This allows the solid deformations to be described using nonlinear soft tissue constitutive laws. Interactions between the Lagrangian and Eulerian fields are achieved by integral transforms with a Dirac delta function kernel $\delta(\mathbf{x})$~\cite{Peskin2002} in Eqs.~\ref{eqn::interact1}~\ref{eqn::interact2}.  For more details of the hybrid IB/FE framework,  please refer to~\cite{Griffith2012-FEIB}.

\subsection{MV-LV Model Construction}
\begin{figure}[h]
	\centering
	\begin{tabular}{ccc}
		\sidesubfloat[]{\includegraphics[height=0.2\textwidth]{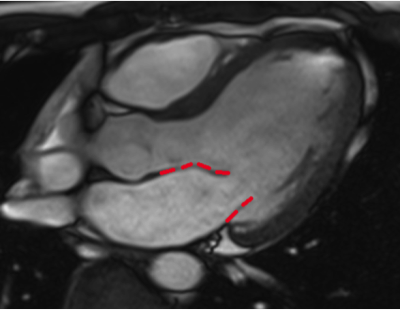}} &
		\sidesubfloat[]{\includegraphics[height=0.2\textwidth]{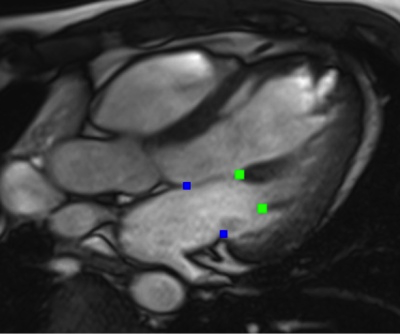}} &
		\sidesubfloat[]{\includegraphics[height=0.22\textwidth]{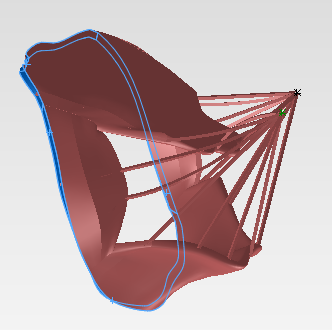}}  \\
		\sidesubfloat[]{\includegraphics[height=0.23\textwidth]{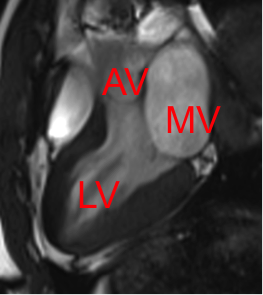}} &
		\sidesubfloat[]{\includegraphics[height=0.23\textwidth]{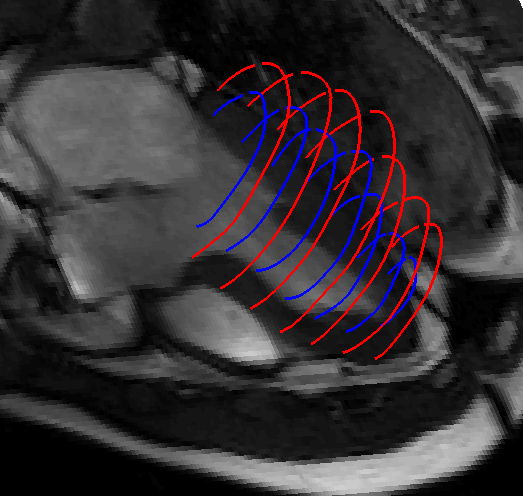}} &
		\sidesubfloat[]{\includegraphics[height=0.25\textwidth]{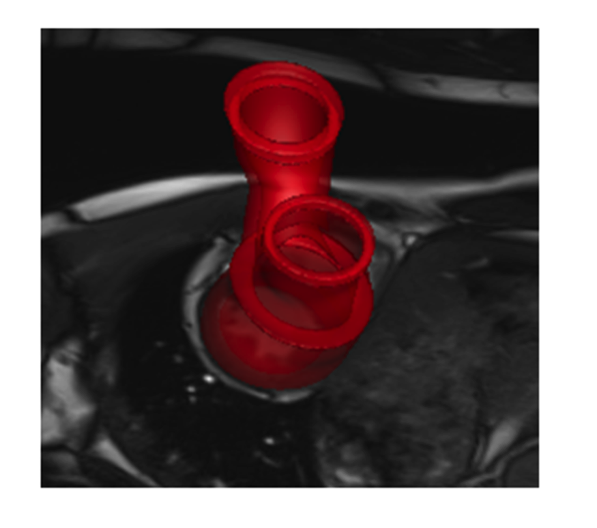}}
	\end{tabular} \\
	\begin{tabular}{cc}
		\sidesubfloat[]{\includegraphics[width=0.25\textwidth]{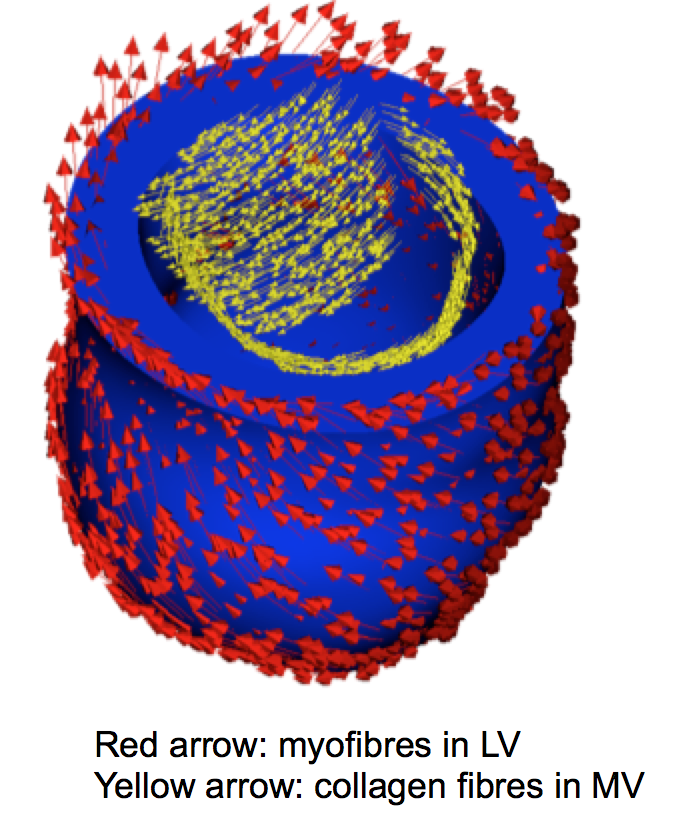}}
		\sidesubfloat[]{\includegraphics[width=0.6\textwidth]{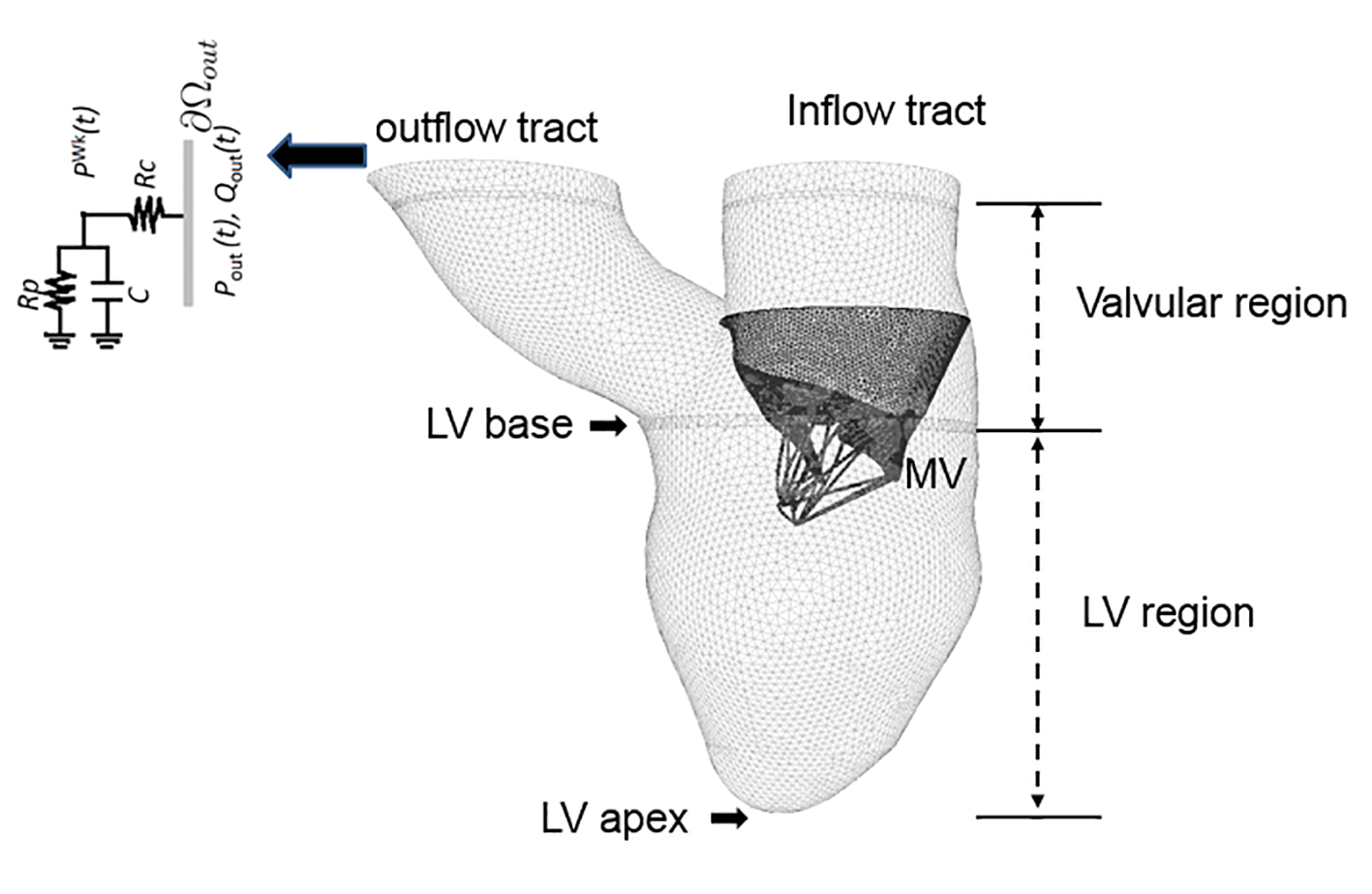}} & 
	\end{tabular}
	
	\caption{The CMR-derived MV-LV model.  (a) The MV leaflets were segmented from a stack
		of MR images of a volunteer at early-diastole, (b) positions of the papillary muscle heads and the annulus ring, (c) reconstructed MV geometry with chordae, (d) a MR image showing the LV and location of the outflow tract (AV) and inflow tract (MV), (e) the LV wall delineation from short and long axis MR images, (f) the reconstructed LV model, in which the LV model is divided into four part: the LV region bellow the LV base, the valvular region, and the inflow and outflow tracts,  (g) the rule-based fibre orientations in the LV and the MV, and  (h) the coupled MV-LV model.}
	\label{f:MV_LV_geo}
\end{figure}

A cardiac magnetic resonance (CMR) study was performed on a healthy volunteer (male, age 28). The study was approved by the local NHS Research Ethics Committee, and written informed consent was obtained before the CMR scan. Twelve imaging planes along the LV outflow tract (LVOT) view were imaged to cover the whole MV region shown in Fig.~\ref{f:MV_LV_geo}(a). LV geometry and function was imaged with conventional short-axis and long-axis cine images. The parameters for the LVOT MV cine images were: slice thickness: 3\,mm with 0 gap, in-plane pixel size: 0.7$\times$0.7\,mm$^2$, field of view: 302 $\times$ 400\,mm$^2$, frame rate: 25 per cardiac cycle. Short-axis cine images covered the LV region from the basal plane to the apex, with slice thickness: 7\,mm with 3\,mm gap, in-plane pixel size: 1.3 $\times$ 1.3\,mm$^2$, and frame rate: 25 per cardiac cycle.  

The MV geometry was reconstructed from LVOT MV cine images at early-diastole, just after the MV opens. The leaflet boundaries were manually delineated from MR images, as shown in Fig.~\ref{f:MV_LV_geo}(a), in which the heads of papillary muscle and the annulus ring were identified as shown in Fig.~\ref{f:MV_LV_geo}(b). The MV geometry and its sub-valvular apparatus were reconstructed using SolidWorks (Dassault Systèmes SolidWorks Corporation, Waltham, MA, USA). Because it is difficult to see the chordal structural in the CMR, we modelled the chordae structure using sixteen evenly distributed chordae tendineae running through the leaflet free edges to the annulus ring, as shown in Fig.~\ref{f:MV_LV_geo}(c), following prior studies~\cite{gao2014finite,ma2012image}. In a similar approach to the MV reconstruction, the LV geometry  was reconstructed from the same volunteer at early-diastole by using both the short-axis and long-axis cine images~\cite{gao2015image,Chen2016}. Fig.~\ref{f:MV_LV_geo}(d) shows the inflow and outflow tracts from one MR image. The LV wall was assembled from the short and long axis MR images (Fig.~\ref{f:MV_LV_geo}(e)) to form the three dimensional reconstruction (Fig.~\ref{f:MV_LV_geo}(f)). The LV model was divided into four regions: the LV and the valvular region and the inflow and the outflow tracts,  as shown in Fig.~\ref{f:MV_LV_geo}(g).

The MV model was mounted into the inflow tract of the LV model according to the relative positions derived from the MR images in Fig.~\ref{f:MV_LV_geo}(g). The left atrium was not reconstructed but modelled as a tubular structure, the gap between the MV annulus ring and the LV model was filled using a housing disc structure. A three-element Windkessel model was attached to the outflow tract of the LV model to provide physiological pressure boundary conditions when the LV is in systolic ejection~\cite{gao2015image}. The chordae were not directly attached to the LV wall since the papillary muscles were not modelled, similar to~\cite{gao2014finite}. The myocardium has a highly layered myofibre architecture, which is usually described using a fibre-sheet-normal ($\f,\s,\n$) system. A rule-based method was used to construct the myofibre orientation within the LV wall. The myofibre angle was assumed to rotate from -60$^o$ to 60$^o$ from endocardium to epicardium, represented by the red arrows in Fig.~\ref{f:MV_LV_geo}(h). In a similar way, the collagen fibres in the MV leaflets were assumed to be circumferentially distributed, parallel along the annulus ring, represented by the yellow arrows in Fig.~\ref{f:MV_LV_geo}(h).

\subsection{Soft Tissue Mechanics}
The total Cauchy stress ($\vsigma$) in the coupled MV-LV system is 
\begin{equation}
\vsigma(\x,t) = \begin{cases}
\sigmaf(\x,t) + \sigmae(\x,t)  & \mbox{for   } \x \in \Omega^\text{s}, \\
\sigmaf(\x,t)       &\mbox{otherwise},
\end{cases} 
\end{equation}
where $\sigmaf$ is the fluid-like stress tensor, defined as 
\begin{equation}
\vsigma^\text{f}(\x,t) = -p\II + \mu[\nabla \u + (\nabla \u)^T].
\end{equation}
$\sigmae$ is the solid stress tensor obtained from the nonlinear soft tissue consitutive laws.  The first Piola-Kirchhoff stress tensor $\PPe$ in Eq.~\ref{eqn::interact1} is related to  $\sigmae$ through 
\begin{equation}
\PPe = J \sigmae \FF^{-T},
\end{equation} 
in which $\FF = \partial {\vec{\chi}}/\partial {\X}$ is the deformation gradient and $J=\det(\FF)$.

In the MV-LV model, we assume the structure below the LV base is contractile (Fig\,\ref{f:MV_LV_geo}(g)), the regions above the LV basal plane, including the MV and its apparatuses, are passive.  Namely, 
\begin{equation}
\PPe = \begin{cases}
\mathbb{P}^\text{p} + \mathbb{P}^\text{a}   &\mbox{below the basal plane}, \\
\mathbb{P}^\text{p}                  &\mbox{above the basal plane}, 
\end{cases}
\end{equation}
where $\mathbb{P}^\text{a}$ and $\mathbb{P}^\text{p}$ are the active and passive Piola-Kirchhoff stress tensors, respectively. 
The MV leaflets are modelled as an incompressible fibre-reinforced material with the strain energy function    
\begin{equation}
W_\text{MV} = C_1 \, (I_1 - 3) + \frac{a_\text{{v}}}{2 b_\text{{v}}} (\exp [b_\text{{v}} ( \max(I_\text{f}^\text{c}, 1) -1)^2] -1),
\label{eqn::mv}
\end{equation} 
in which $I_1 = \text{trace}(\CC)$ is the first invariant of the right Cauchy-Green deformation tensor $\CC = \FF^T \FF$, $I_f^\text{c} = f_0^\text{c} \cdot (\CC f_0^\text{c})$ is the squared stretch along the collagen fibre direction, and $f_0^\text{c}$ denotes the collagen fibre orientation in the reference configuration. The $\max()$ function ensures the embedded collagen network only bears the loads when stretched, but not in compression. $C_1$, $a_\text{v}$, and $b_\text{v}$ are material parameters adopted from a prior study~\cite{gao2014finite} and listed in Table \ref{tab::passiveMat}. The passive stress tensor $\mathbb{P}^\text{p}$ in the MV leaflets is
\begin{equation}
\mathbb{P}^\text{p} = \D{W_\text{MV}}{\FF} - C_1 \FF^{-T} + \beta_\text{s} \log(I_3) \FF^{-T},
\label{eqn::stressMV}
\end{equation}
where $I_3 = \det(\CC)$, and $\beta_\text{s}$ is the  bulk modulus for ensuring the incompressibility of immersed solid, so that the pressure-like term $C_1 \FF^{-T}$ ensures the elastic stress response is zero when $\FF = \II$. 

We model the chordae tendineae as the Neo-Hookean material,  
\begin{equation}
W_\text{chordae} = C\, (I_1 - 3),
\label{eqn::chordae}
\end{equation}
where $C$ is the shear modulus. We further assume $C$ is much larger in systole when the MV is closed than in diastole when the valve is opened. The much larger value of $C$ models the effects of  papillary muscle contraction. Values of $C$ are listed in Table \ref{tab::passiveMat}.  $\mathbb{P}^\text{p}$ for the chordae tendineae is similarly derived as in Eq.~\ref{eqn::stressMV}.  

The passive response of the LV myocardium is described using the Holzapfel-Ogden model~\cite{holzapfel2009constitutive},   
\begin{equation}
\begin{split}
W_\text{myo} = &\frac{a}{2b} \exp[b(I_1 - 3)] + \sum_{i=f,s}\frac{a_i}{2b_i}\{ \exp [b_i ( \max(I_{4i},1) -1)^2] -1 \}  \\
&+ \frac{a_\text{fs}}{2b_\text{fs}} \{ \exp[b_\text{fs} (I_\text{8fs})^2] -1 \}
\end{split}
\label{eqn::myostrainenergy}
\end{equation}
in which $a,b, a_\text{f}, b_\text{f}, a_\text{s}, b_\text{s}, a_\text{fs}, b_\text{fs}$ are the material parameters, $I_\text{{4f}}$, $I_\text{{4s}}$ and $I_\text{8fs}$ are the strain invariants related to the the myofibre orientations. Denoting the myofibre direction in the reference state is $\f_0$ and the sheet direction is $\s_0$,  we have 
\begin{equation}
I_\text{4f} = \f_0 \cdot (\CC \f_0),\, I_\text{4s} = \s_0 \cdot (\CC \s_0), \mbox{and }\, I_\text{8fs} = \f_0\cdot (\CC \s_0). 
\end{equation}  

The myocardial active stress is defined as 
\begin{equation}
\mathbb{P}^\text{a} = J \, T \, \FF \,\f_0 \otimes \f_0
\label{eqn::stressAct}
\end{equation} 
where $T$ is the active tension described by the myofilament model of Niederer et al.~\cite{niederer2006quantitative}, using a set of ordinary differential equations involving the intracellular calcium transient ($\Ca^{2+}$), sarcomere length and the active tension at the resting sarcomere length ($T^\text{req}$). In our simulations, we use the same parameters as in ref.~\cite{niederer2006quantitative}, except that $T^\text{req}$ is adjusted to yield realistic contraction as the imaged volunteer.   

All the constitutive parameters in Eqs.\ref{eqn::mv}, \ref{eqn::chordae}, \ref{eqn::myostrainenergy} are summarized in Table\,\ref{tab::passiveMat}.

\subsection{Boundary Conditions and Model Implementation}
Because only the myocardium below the LV basal plane contracts, we fix the LV basal plane along the circumferential and longitudinal displacements, but allow the radial expansion. The myocardium below the LV basal plane is left free to move. The valvular region is assumed to be much softer than the LV region.  In diastole, a maximum displacement of 6~mm is allowed in the valvular region using a tethering force. In systole, the valve region is gradually pulled back to the original position. The inflow and outflow tracts are fixed. Because the MV annulus ring are attached to a housing structure which is fixed, no additional boundary conditions are applied to the MV annulus ring. Fluid boundary conditions are applied to the top planes of the inflow and outflow tracts. The function of the aortic valve is modelled simply:  the aortic valve is either fully opened or fully closed, determined by the pressure difference between the values inside the LV chamber and the aorta. After end-diastole, the LV region will contract simultaneously triggered by a spatially homogeneously prescribed intracellular $\Ca^{2+}$ transient \cite{Chen2016}, as shown in Fig.~\ref{fig::averageProfiles}. The flow boundary conditions in a cardiac cycle are summarized below. 
\begin{itemize}
	\item \textbf{Diastolic filling:} A linearly ramped pressure from 0 to a population-based end-diastolic pressure  (EDP=8\,mmHg) is applied to the inflow tract over 0.8\,s, which is slightly longer than the actual diastolic duration of the imaged volunteer (0.6 s).  In diastole about 80\% of diastolic filling volume is due to the sucking effect of the left ventricle in early-diastole \cite{nishimura1997evaluation}. This negative pressure field inside the LV cavity is due to the myocardial relaxation.  We model this sucking effect using an additional pressure loading applied to the endocardial surface, denoted as  $P_\text{endo}$, which is linearly ramped from 0 to 12\,mmHg over 0.4\,s, and then linearly decreased to zero at end-diastole. The value of $P_\text{endo}$ is chosen by matching the simulated end-diastolic volume to the measured data from CMR images.  Blood flow is not allowed to move out of the LV cavity through the inflow tract in diastole. Zero flow boundary conditions are applied to the top plane of the outflow tract.  
	\item \textbf{Iso-volumetric contraction:} Along the top plane of the inflow tract, the EDP loading is maintained, but we allow free fluid flow in and out of the inflow tract. Zero flow boundary conditions are retained for the outflow tract. The duration of the iso-volumetric contraction is determined by the myocardial contraction and ends when the aortic valve opens. The aortic valve opens when the LV pressure is higher than the pressure in the aorta, which is initially set to be the cuff-measured diastolic pressure in the brachial artery, 85\,mmHg.
	\item \textbf{Systolic ejection:} When the aortic valve opens, a three-element Windkessl model is coupled to the top plane of the outflow tract to provide afterload. The volumetric flow rates across the top plane of the outflow tract is calculated from the three-dimensional MV-LV model, and fed into the Windkessel model \cite{griffith2012immersed}, which returns an updated pressure for the outflow tract in the next time step. The systolic ejection phase ends when the left ventricle cannot pump any flow through the outflow tract, and the Windkessel model is detached. 
	\item \textbf{Iso-volumetric relaxation:} Zero flow boundary conditions are applied to both the top planes of the outflow and inflow tracts until the total cycle ends at 1.2 seconds.   
\end{itemize} 

The coupled MV-LV model is immersed in a 17cm $\times$ 16cm $\times$ 16cm fluid box. A basic time step size $\Delta t_0=1.22\times10^{-4}$\,s is used in the diastolic and relaxation phases, a reduced time step size (0.25\,$\Delta t_0$) is used in the early systole with a duration of 0.1\,s, and an even smaller time step of 0.125\,$\Delta t_0$ is used in the remainder of the systolic phase.  Because explicit time stepping is used in the numerical simulations \cite{Griffith2012-FEIB}, we need to use a time step size small enough to avoid numerical instabilities,  particularly during the systolic phase to resolve the highly dynamic LV deformation.  The MV-LV model is implemented using the open-source IBAMR software framework (\url{https://github.com/IBAMR/IBAMR}), which provides an adaptive and distributed-memory parallel implementation of the IB methods. 

\section{Results}

\begin{figure}[h]
	\includegraphics[width=0.9\textwidth]{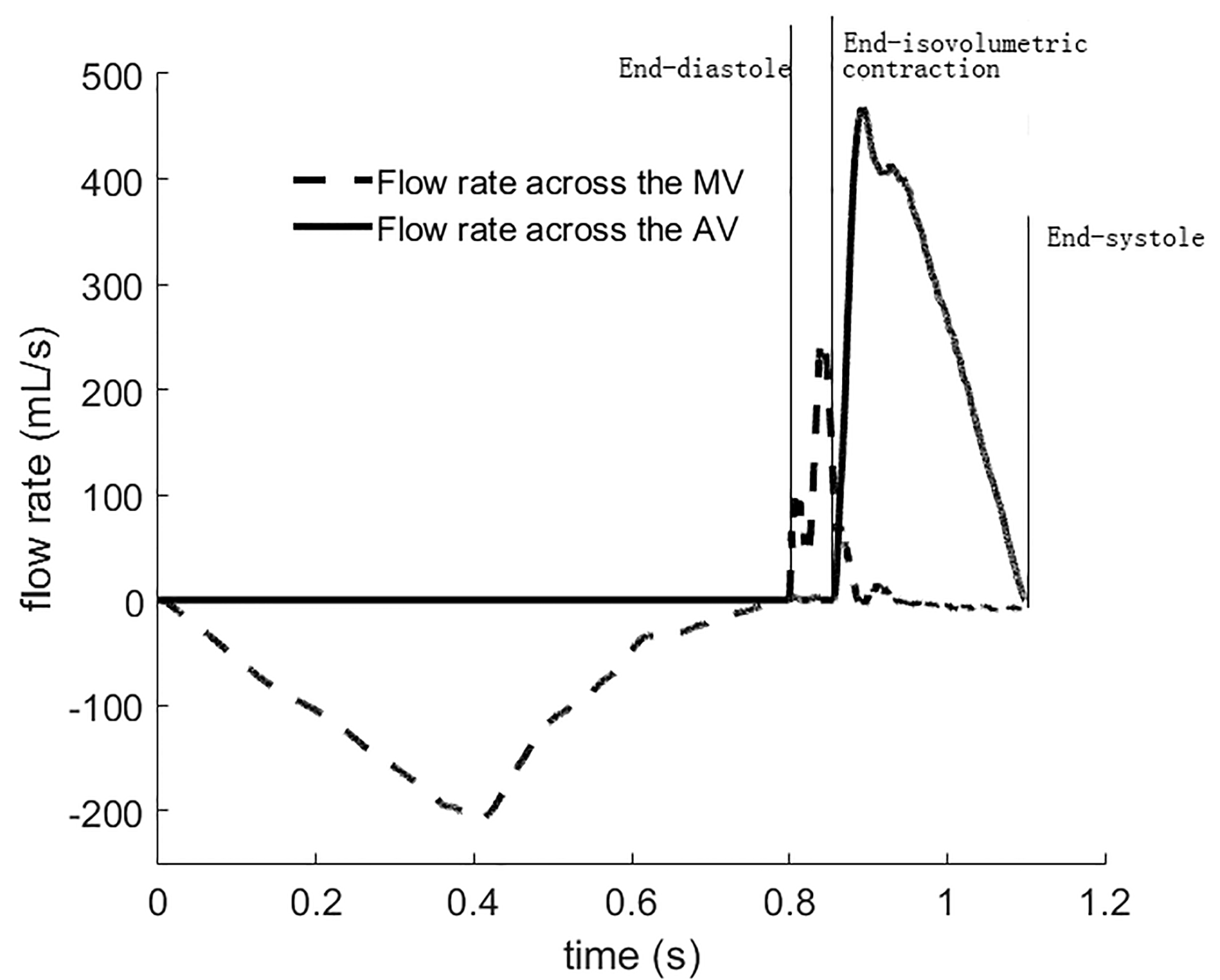}
	\caption{Flow rates across the AV and MV from diastole to systole. Diastolic phase: 0\,s to 0.8\,s; Systolic phase: 0.8\,s and onwards. Positive flow rate means the blood flows out of the LV chamber.}
	\label{fig::avmvflowrates}
\end{figure}

Fig.~\ref{fig::avmvflowrates} shows the computed volumetric flow rates across the MV and the AV from beginning of diastole to end-systole. In diastole, the volumetric flow rate across the MV linearly increases with $P_\text{endo}$, with a maximum value of 210~mL/s at 0.4\,s.  Diastolic filling is maintained by the increased pressure in the inflow tract, but with decreased flow rates until end of diastole at 0.8\,s. The negative flow rate in Fig.\,\ref{fig::avmvflowrates} indicates the flow is entering the LV chamber. After end-diastole, the myocardium starts to contract, and the central LV pressure increases until it exceeds the aortic pressure (initially set to be 85\,mmHg) at 0.857\,s.  During iso-volumetric contraction, the MV closes with a total closure regurgitation flow of 7.2\,mL, around 10\% of the total filling volume, which is comparable to the value reported by Laniado et al.~\cite{Laniado104}. There is only minor regurgitation across the MV during systolic ejection after the iso-volumetric contraction phase. Blood is then  ejected out of the ventricle through the AV, and the flow rate across the AV during systole reaches a peak value of 468\,mL/s (Fig.~\ref{fig::avmvflowrates}). The total ejection duration is 243\,ms with a stroke volume of 63.2\,mL. The total blood ejected out of the LV chamber, including the regurgitation across the MV, is 72.1\,mL, which corresponds to an ejection fraction of 51\%.

\begin{figure}[h]
	\includegraphics[width=1.0\textwidth]{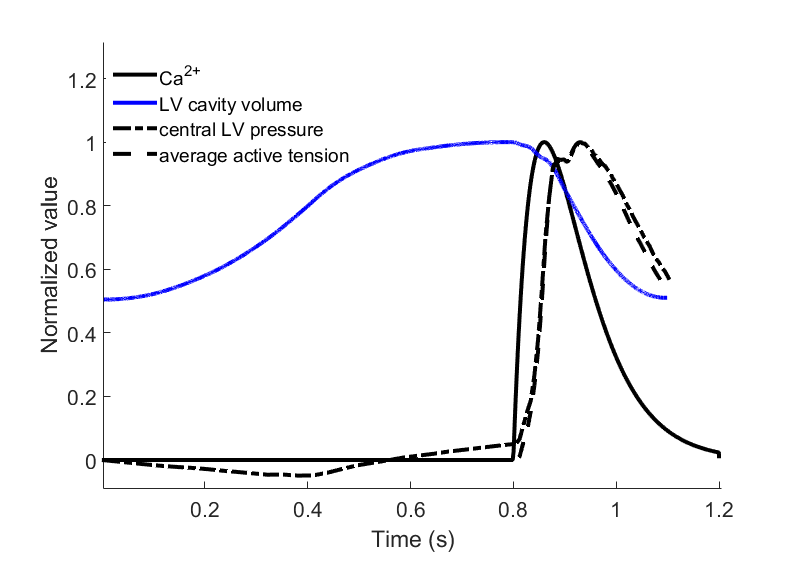}
	\caption{Normalized intracellular \Ca$^2+$, LV cavity volume, central LV pressure and average myocardial active tension. All curves are normalized to their own maximum values, which are: 1$\mu$Mol for \Ca$^2+$, 145\,mL for LV cavity volume, 162\,mmHg for central LV pressure, 96.3\,kPa for average myocardial active tension.}
	\label{fig::averageProfiles}
\end{figure}

Fig.~\ref{fig::averageProfiles} shows the profiles of the normalized intracellular \Ca$^{2+}$, LV cavity volume, central LV pressure, and the average myocardial active tension from diastole to systole. Until mid-diastole (0\,s to 0.56\,s), the central LV pressure is negative, and the associated diastolic filling volume is around 65\,mL, which is 90\% of the total diastolic filling volume. In late-diastole, the LV pressure becomes positive. There is a delay between the myocardial active tension and the intracellular \Ca$^{2+}$ profile, but the central LV pressure  follows the active tension closely throughout the cycle as shown in Fig~\ref{fig::averageProfiles}. 

\begin{figure}[h]
	\begin{tabular}{cc}
		\sidesubfloat[]{\includegraphics[width=0.3\textwidth]{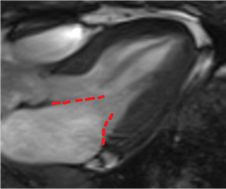}} &
		\sidesubfloat{\includegraphics[width=0.4\textwidth]{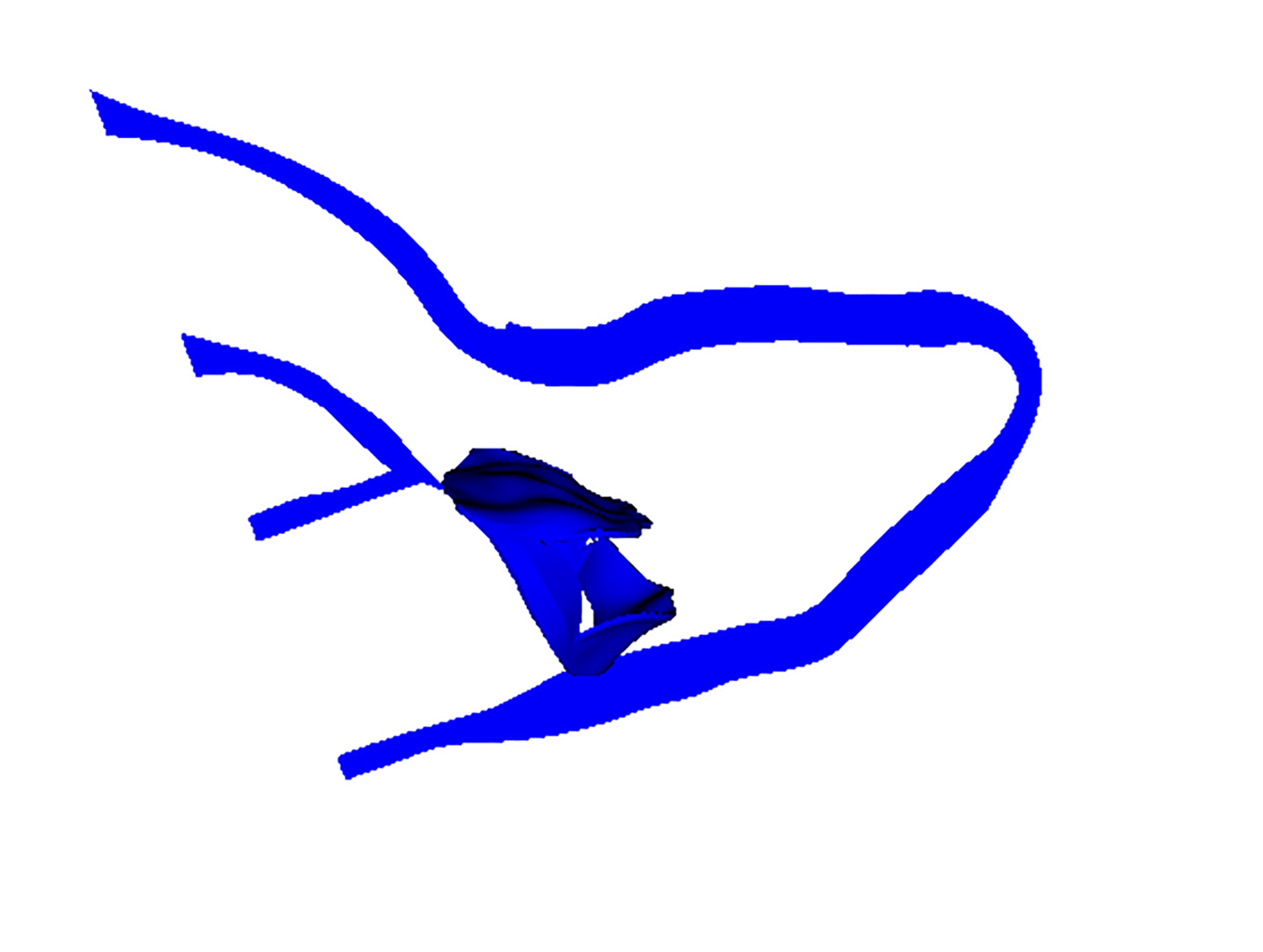}}  \\
		  \addtocounter{subfigure}{-1}
		\sidesubfloat[]{\includegraphics[width=0.3\textwidth]{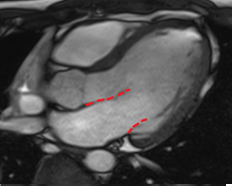}} &
		\sidesubfloat{\includegraphics[width=0.4\textwidth]{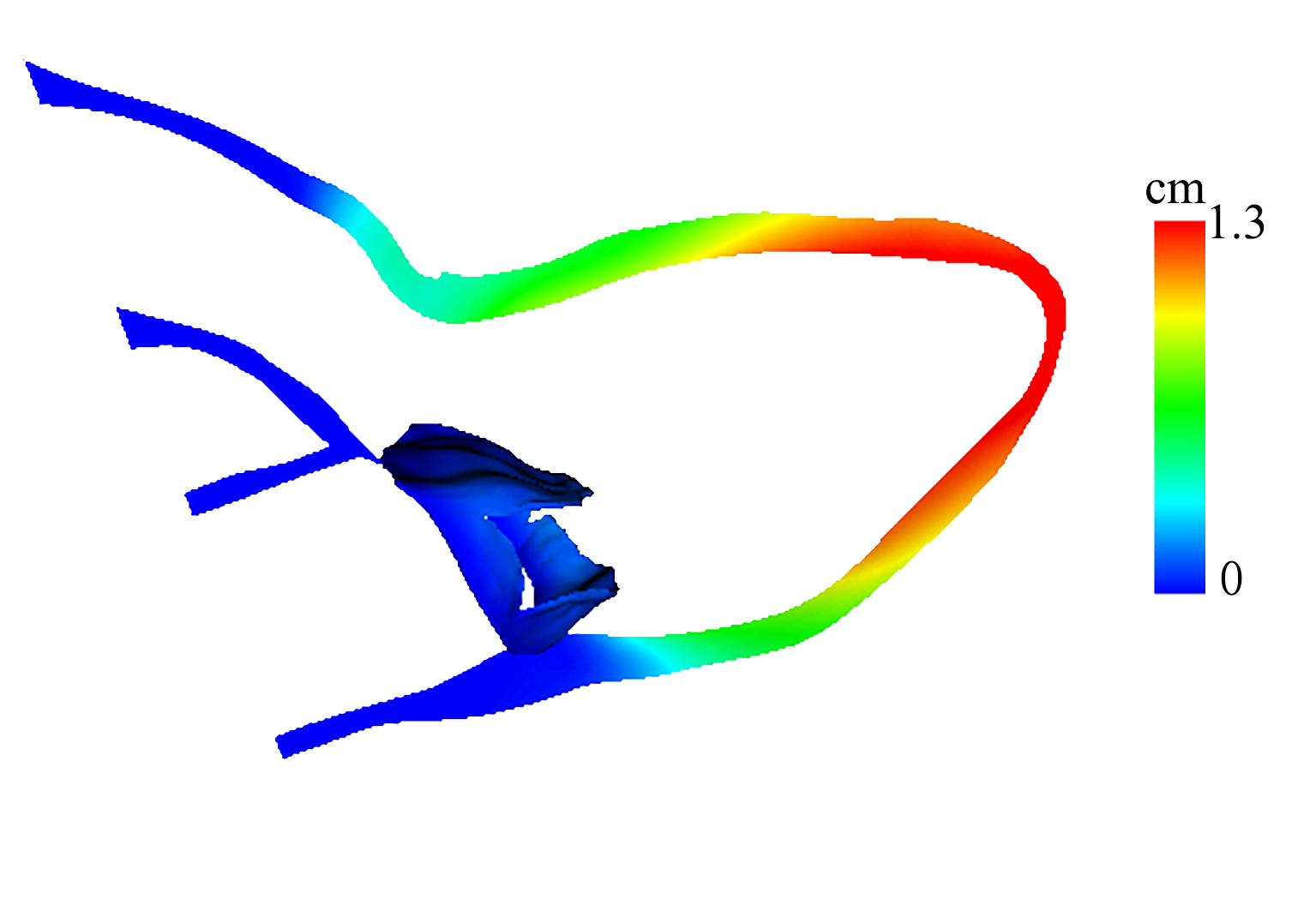}}  \\
				  \addtocounter{subfigure}{-1}
		\sidesubfloat[]{\includegraphics[width=0.3\textwidth]{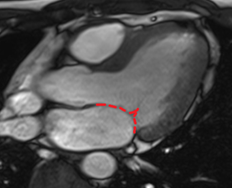}} &
		\sidesubfloat{\includegraphics[width=0.4\textwidth]{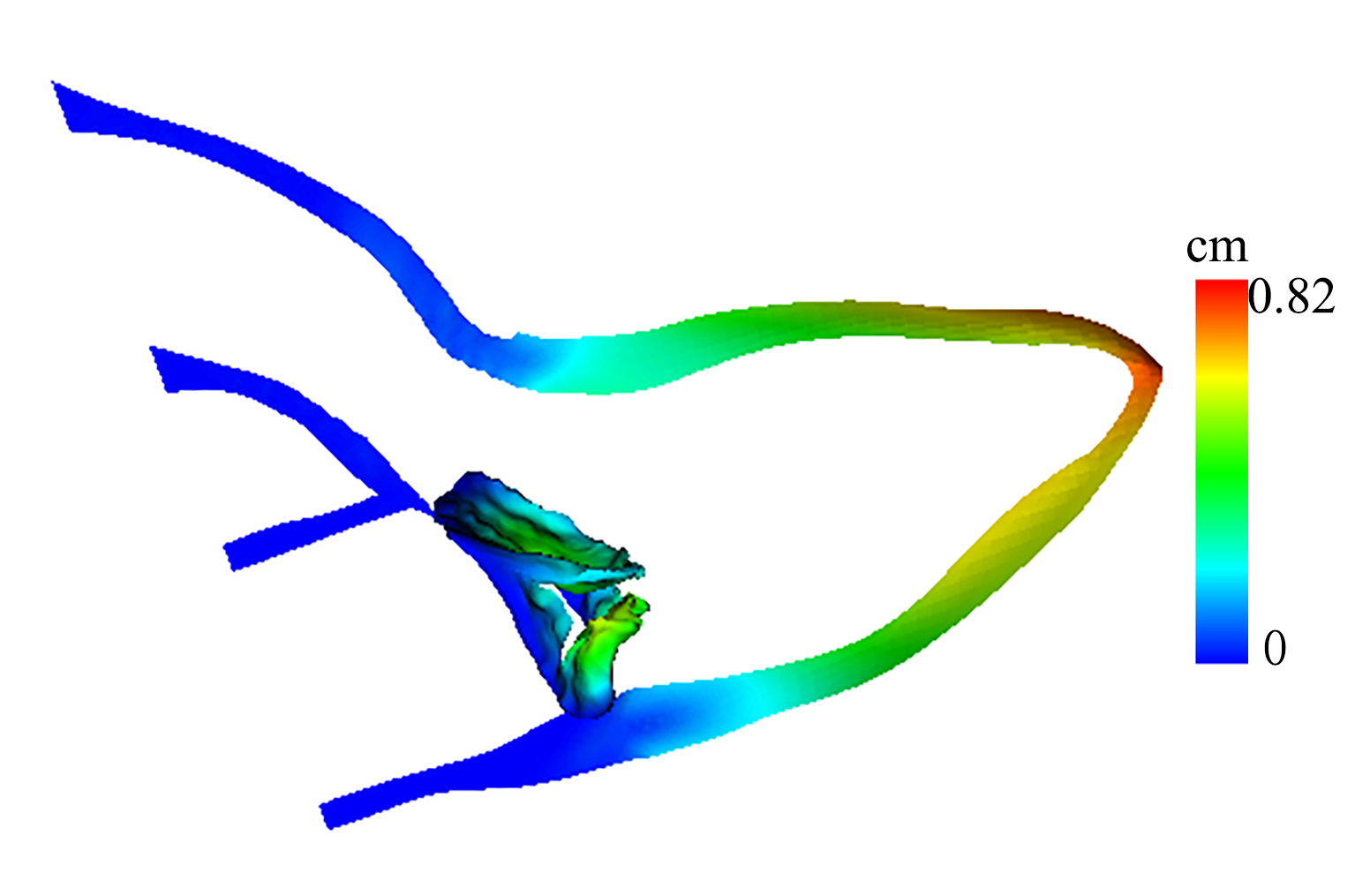}}  \\
	\end{tabular}
	\caption{Comparisons between the MV and LV structures at (a) reference configuration, (b) end-diastole,  and (c) end-systole, and the corresponding CMR cine images (left). Coloured by the displacement magnitude.}
	\label{fig::LV-MV-geo-MRI}
\end{figure}

Fig.~\ref{fig::LV-MV-geo-MRI} shows the deformed MV leaflets along with the corresponding CMR cine images at early-diastole (the reference state), end-diastole, and mid-systole. In general, the in vivo MV and LV dynamics from diastole to systole are qualitatively captured well by the coupled MV-LV model.    However,  a discrepancy is observed during the diastolic filling, when the MV orifice in the model is not opened as widely as in the CMR cine image (Fig.~\ref{fig::LV-MV-geo-MRI}(b)). In addition, the modelled MV leaflets have small gaps near the commissure areas even in the fully closure state.  This is partially caused by the finite size of the regularized delta function at the interface and uncertainties in MV geometry reconstruction using CMR images.   
   
\begin{figure}
	\begin{tabular}{cc}
		\sidesubfloat[]{\includegraphics[height=0.5\textwidth]{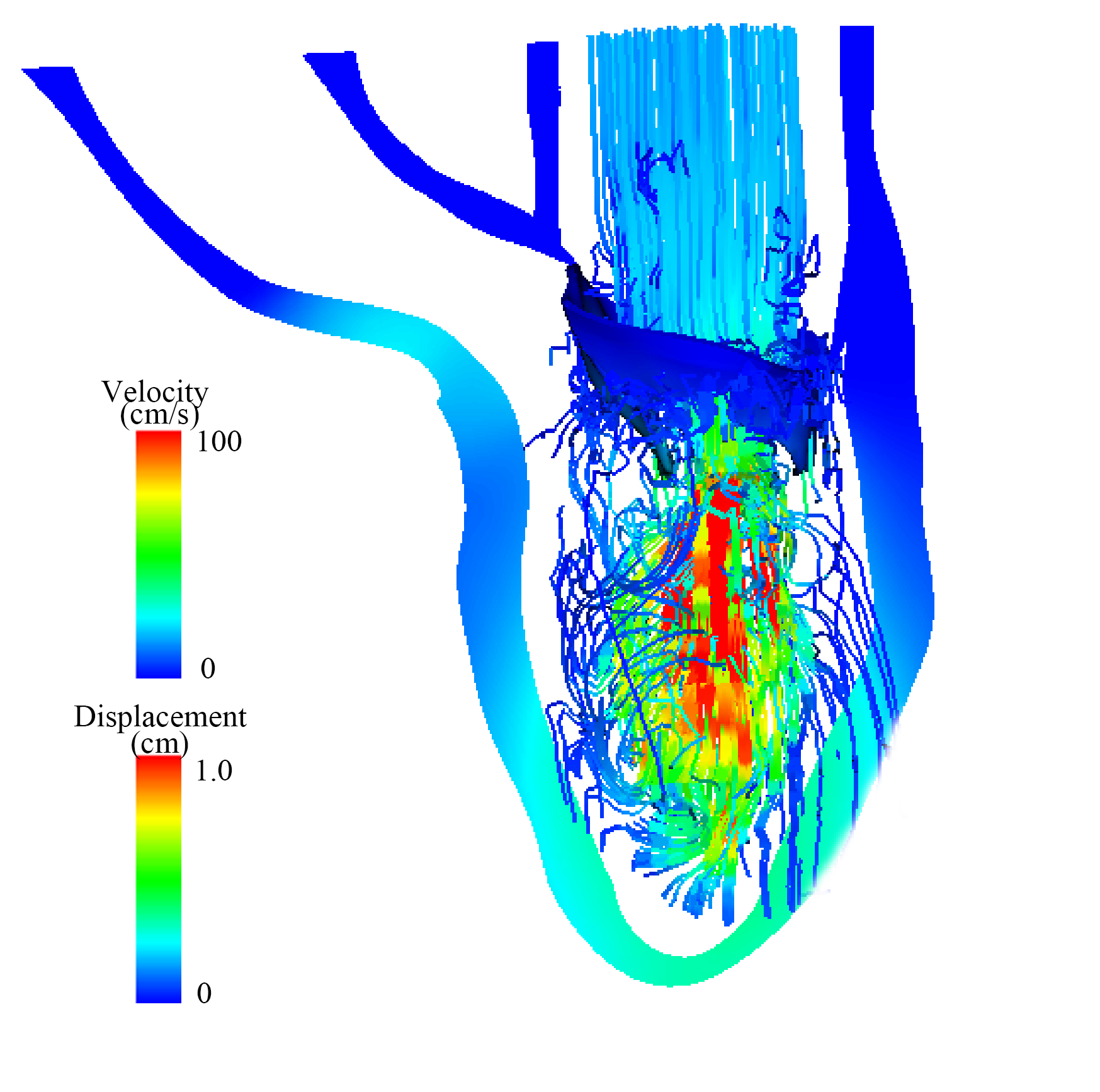}} &
		\sidesubfloat[]{\includegraphics[height=0.5\textwidth]{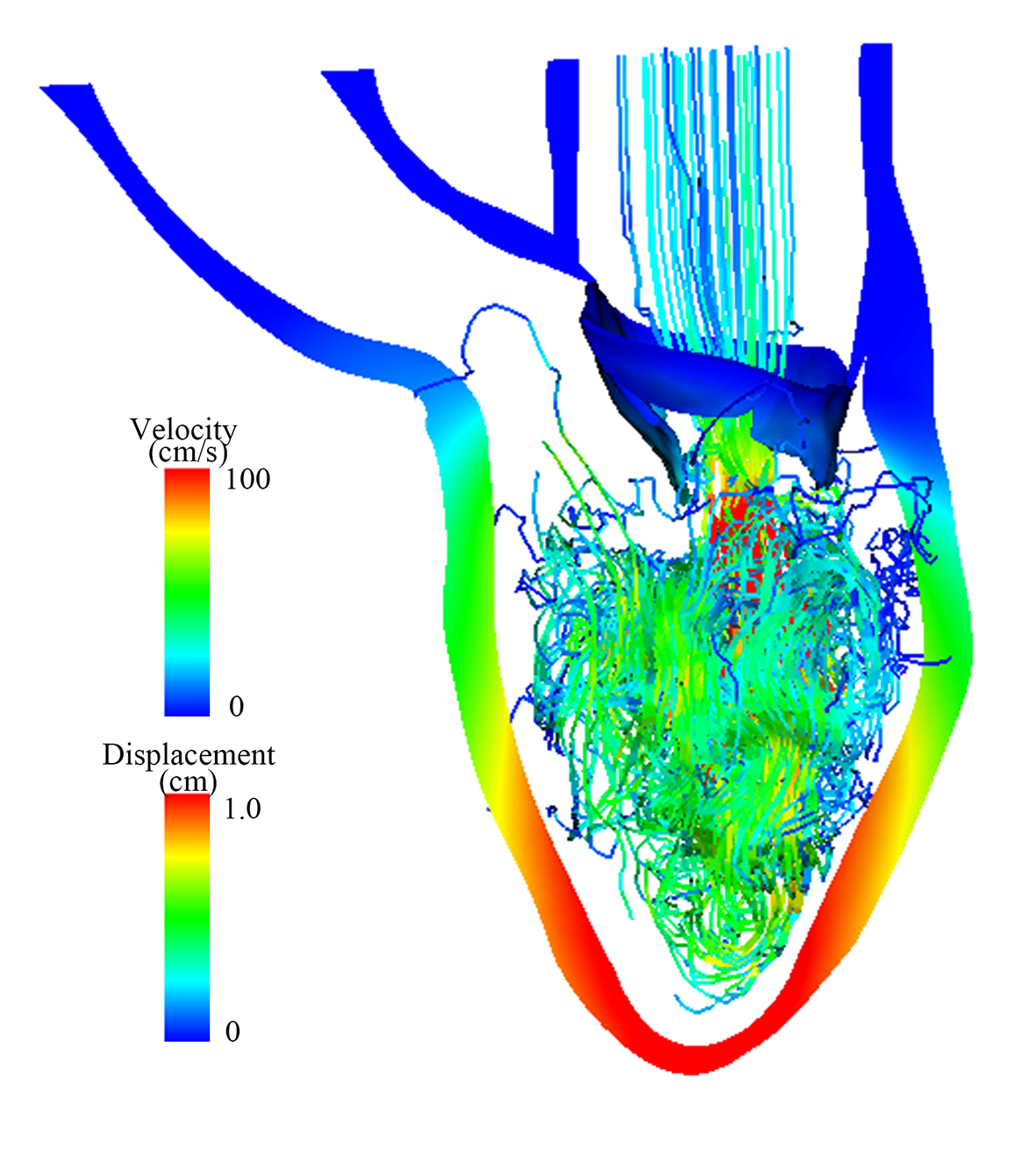}}  \\
		\sidesubfloat[]{\includegraphics[height=0.5\textwidth]{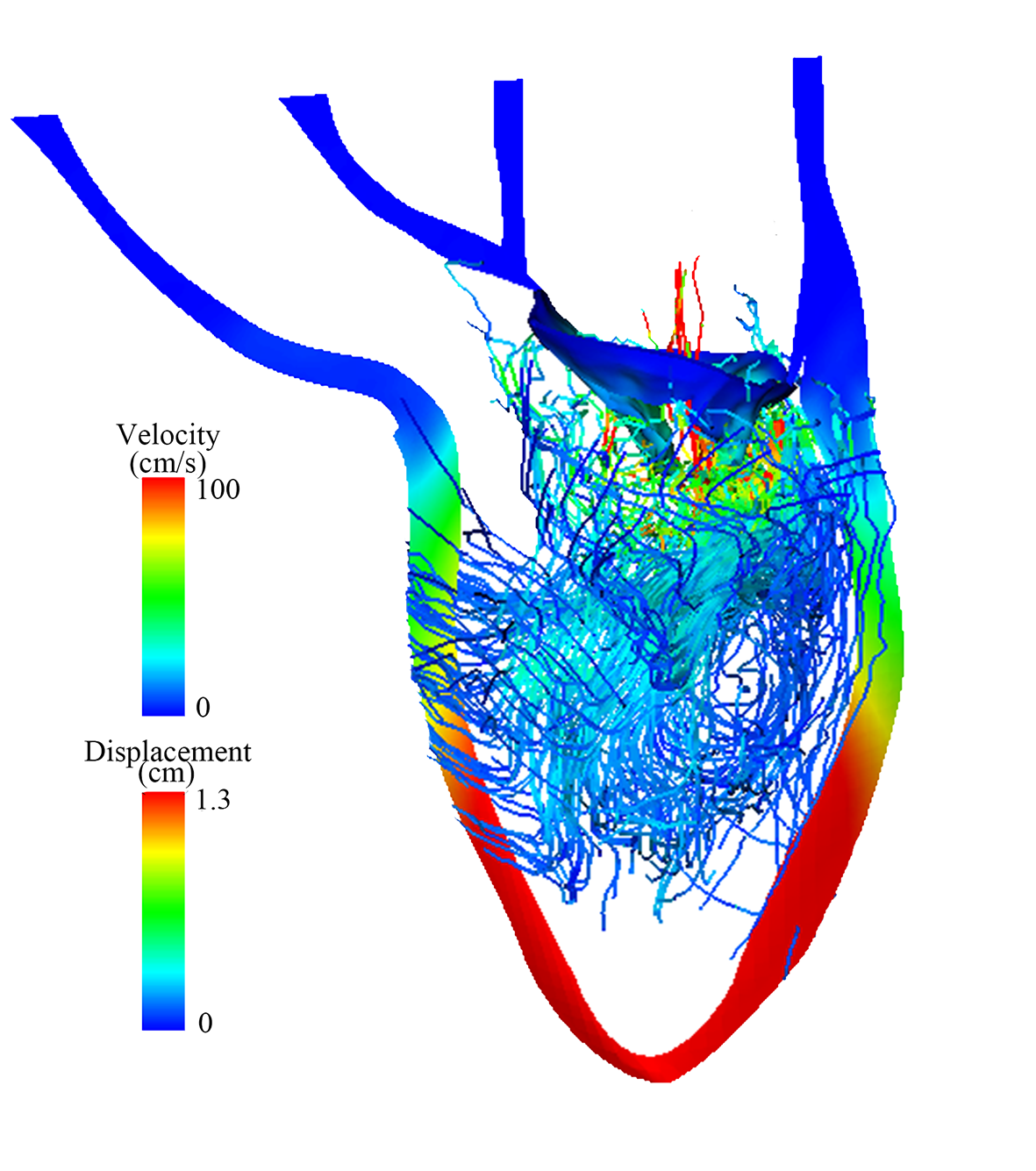}} &
		\sidesubfloat[]{\includegraphics[height=0.5\textwidth]{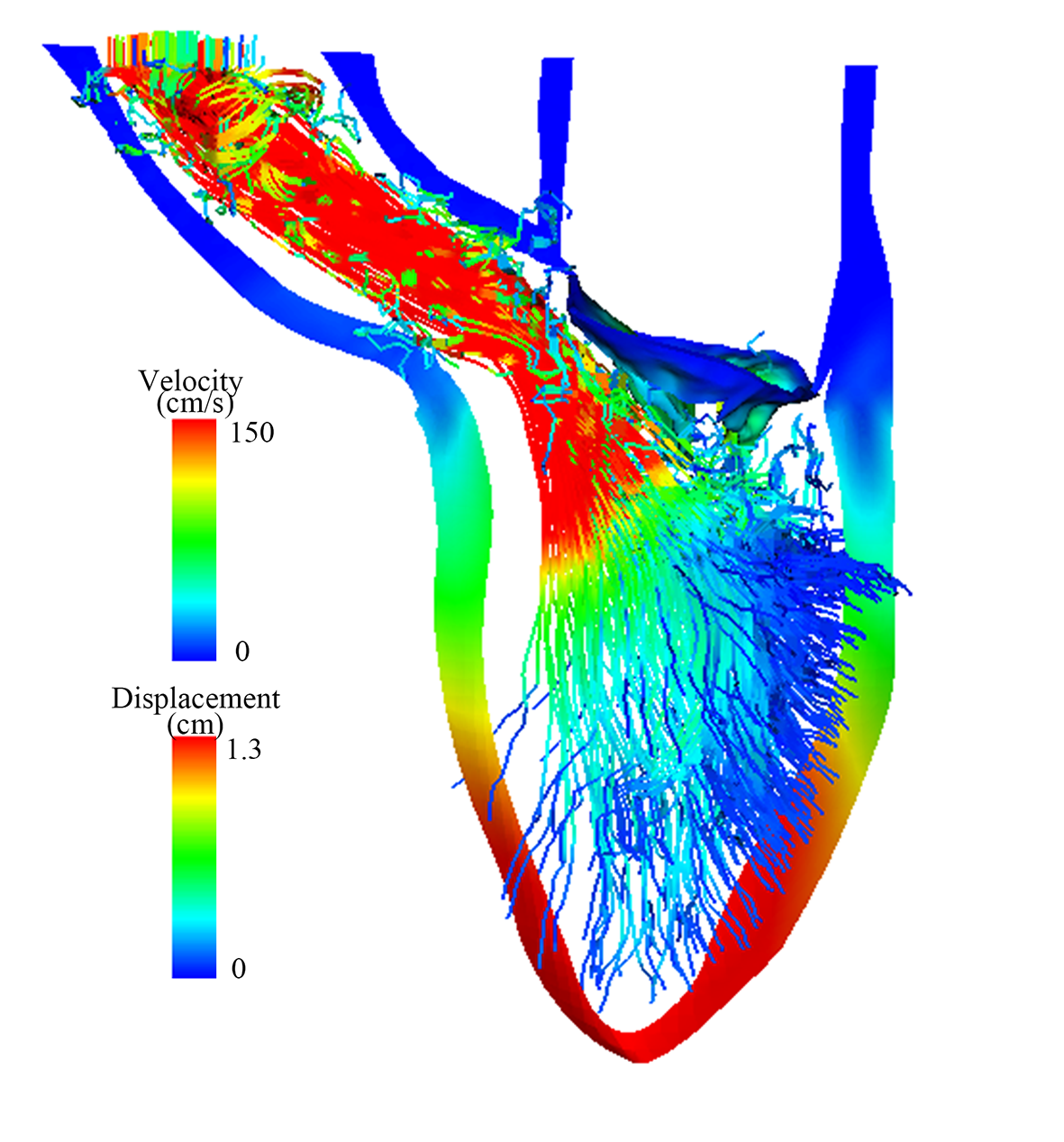}}  \\
	\end{tabular}
	\caption{Streamlines in the MV-LV model at early-diastolic filling (a), late-diastolic filling (b), when isovolumtric contraction ends (c), and at the mid-systole. Streamline are colored by velocity magnitude, the LV wall and MV are colored by the displacement magnitude. Red : high; blue: low }
	\label{fig::streamlines}
\end{figure}

Figs.~\ref{fig::streamlines}(a, b, c, d) show the streamlines at early-diastolic filling, late-diastolic filling, when the MV is closing (iso-volumtric contraction), and mid-systolic ejection when the left ventricle is ejecting. During the diastolic filling (Fig.~\ref{fig::streamlines}(a)), the blood flows directly through the MV into the LV chamber towards the LV apex, in late-diastole in Fig.~\ref{fig::streamlines}(b), the flow pattern becomes highly complex. When iso-volmeric contraction ends, the MV is pushed back towards the left atrium. In mid-systole, the blood is pumped out of the LV chamber through the aortic valve into the systemic circulation, forming a strong jet as shown in  Fig.~\ref{fig::streamlines}\,(d).        
   
\begin{figure}
	\sidesubfloat[]{\includegraphics[width=0.4\textwidth]{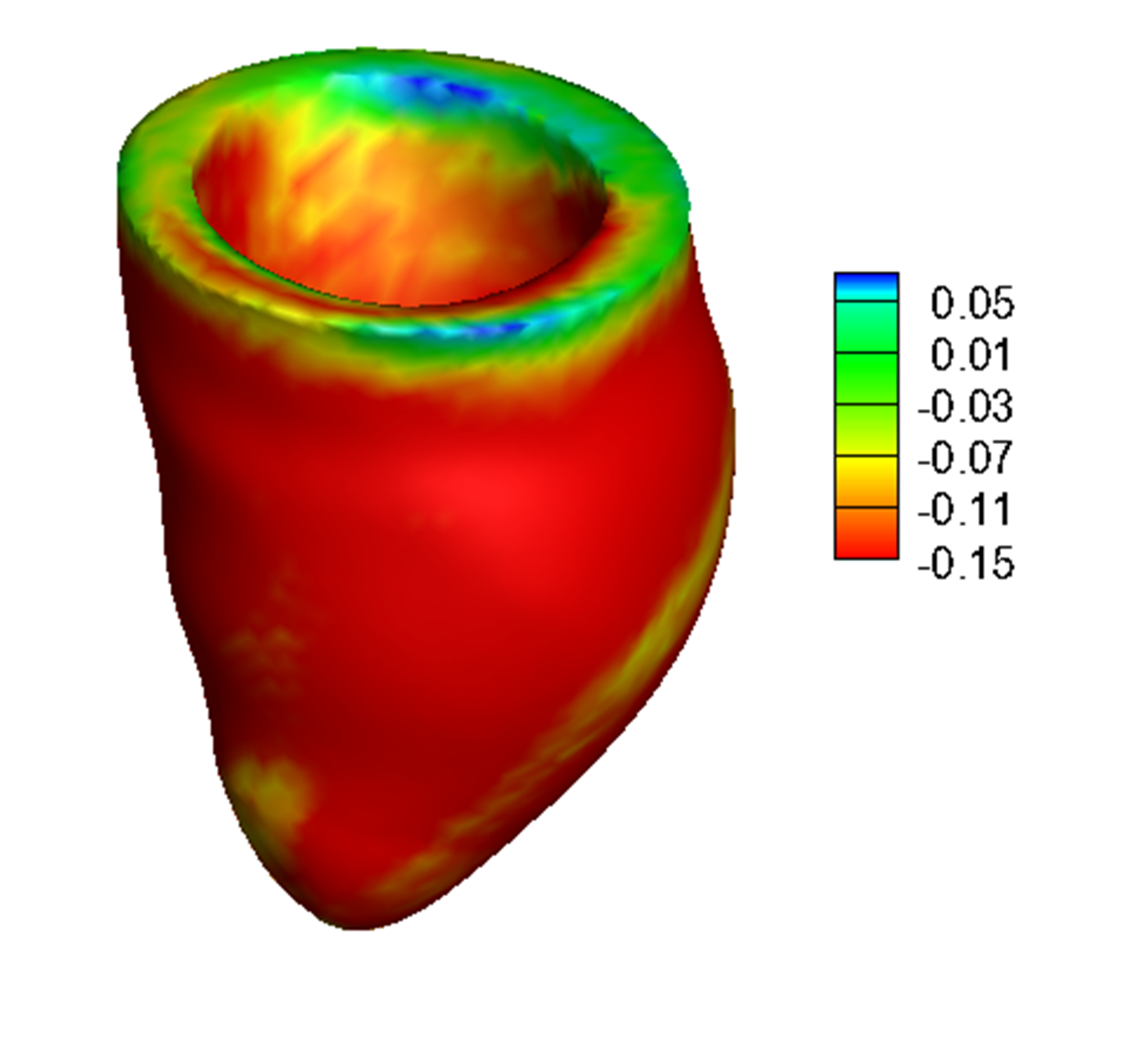}} \\
	\begin{tabular}{cc}	
		\sidesubfloat[]{\includegraphics[width=0.45\textwidth]{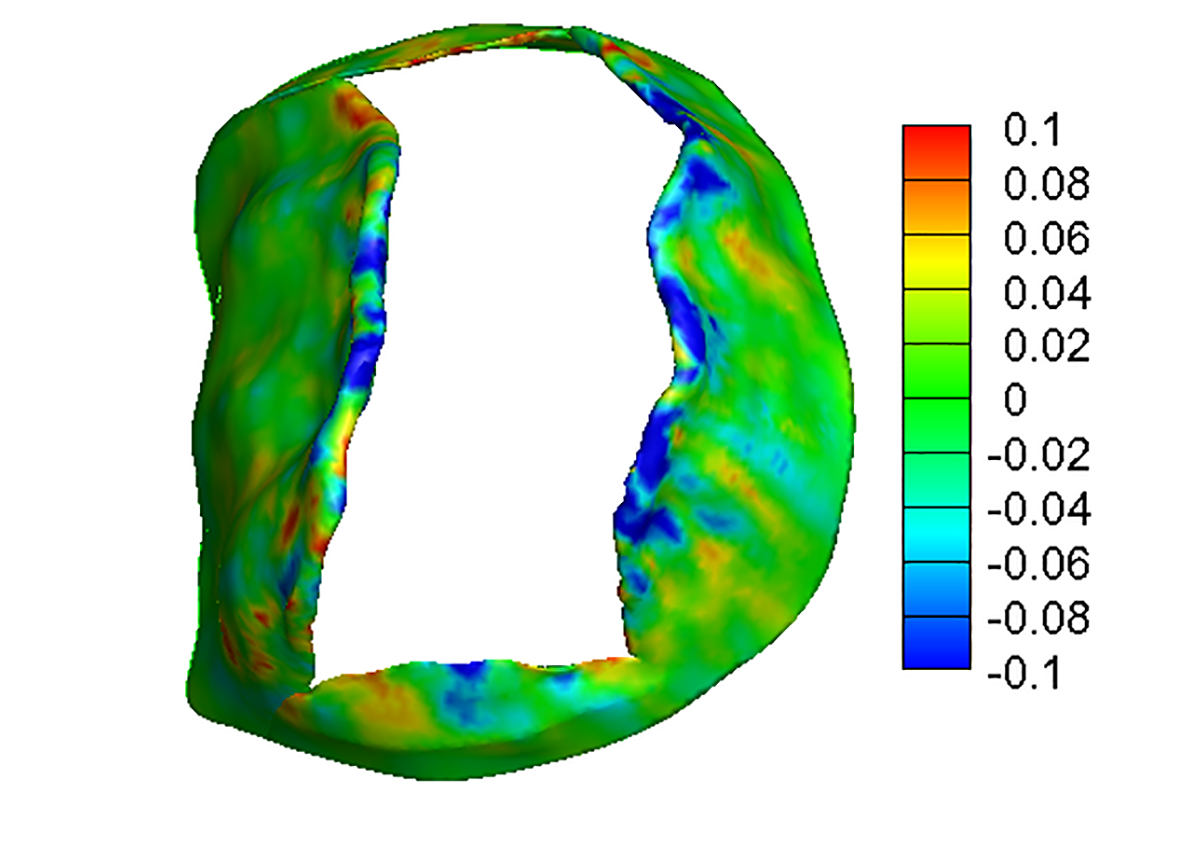}} &	
		\sidesubfloat[]{\includegraphics[width=0.45\textwidth]{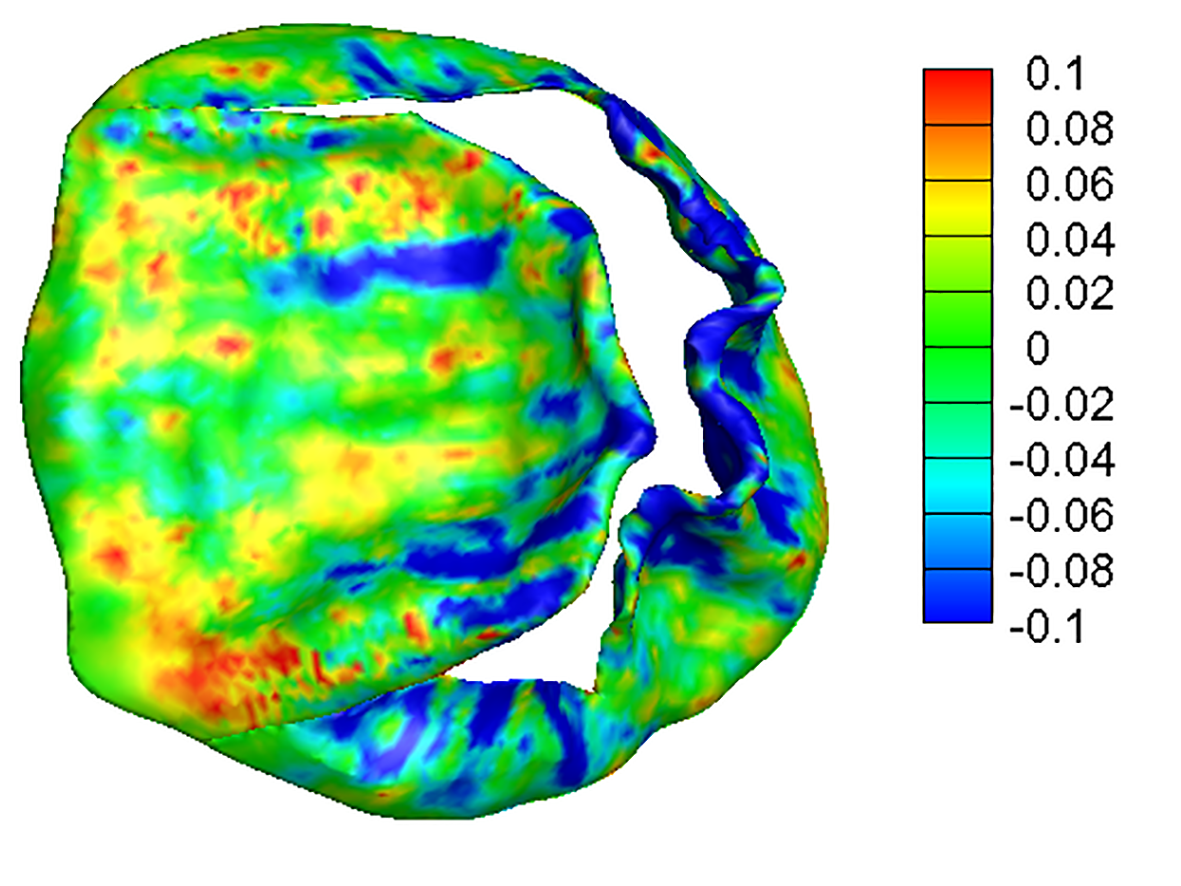}} \\
	\end{tabular}
	\caption{Distributions of fibre strain in the left ventricle at end-systole (a), in the MV at end-diastole (b) and end-systole (c).}
	\label{fig::fibrestrain}
\end{figure}

The LV systolic strain related to end-diastole is shown in Fig.~\ref{fig::fibrestrain} (a), which is negative throughout most of the region except near the basal plane,  where the LV motion is artificially constrained in the model. The average myocardial strain along myofibre direction is -0.162$\pm$0.05.  Fig.~\ref{fig::fibrestrain}(b) is the fibre strain in the MV leaflets at end-diastole, the leaflets are mostly slightly stretched during the diastolic filling. In systole, because of the much higher pressure in the LV, the leaflets are pushed towards the left atrium side as shown in Fig.\,\ref{fig::fibrestrain}(c). Near the leaflet tip and the commissiour areas, the leaflets are highly compressed, while in the trigons near the annulus ring, the leaflet is stretched.

   
From Fig.\,\ref{fig::averageProfiles}, one can see that the applied endocardial pressure ($P_\text{endo}$) creates a negative pressure inside the LV chamber, similar to the effects of the myocardial active relaxation.  We further investigate how $P_\text{endo}$ affects the MV-LV dynamics by varying its value from 8\,mmHg to 16\,mmHg, and the effects without $P_\text{endo}$ but with an increased EDP from 8\,mmHg to 20\,mmHg.  We observe that with an increased $P_\text{endo}$, the peak flow rate across the MV during the filling phase becomes higher with more ejected volume through the aortic valve.  We also have a longer ejection duration, shorter iso-volumetric contraction time, and higher ejection fraction as a result of increasing $P_\text{endo}$.  On the other hand, if we don't apply $P_\text{endo}$, a much greater and nonphysiological EDP is needed for the required ejection fraction. For example, with EDP=8\,mmHg, the ejection fraction is only 29\%.  Only when EDP=20\,mmHg, the pump function is comparable to the case with EDP=8\,mmHg and $P_\text{endo}$ = 16\,mmHg.  These results are summarized in Table \ref{tab::EDPEPLEffects}.

\section{Discussion}
This study demonstrates the feasibility of integrating a MV model with a LV model from a healthy volunteer based on in vivo CMR images. This is the first physiologically based MV-LV model with fluid structure interaction that includes nonlinear hyperelastic constitutive modelling of the soft tissue.  The coupled MV-LV model is used to simulate MV dynamics, LV wall deformation, myocardial active contraction, as well as intraventricular flow. The modelling results are in reasonable quantitative agreement with in vivo measurements and clinical observations. For example, the peak aortic flow rate is 468\,mL/s, close to the measured peak value (498\,mL/s); the ejection duration is 243\,ms, and the measured value is around 300\,ms; the peak LV pressure is 162\,mmHg, comparable to the cuff-measured peak blood pressure 150\,mmHg; the average LV systolic strain is around -0.16, which also lies in the normal range of healthy subjects \cite{Mangion2016Estimating}. 

Diastolic heart failure is usually associated with impaired myocardial relaxation and increased filling pressure~\cite{hay2005role,zile2004diastolic}.    In this study, we model the effects of myocardial relaxation by applying an endocardial surface pressure $P_\text{endo}$.  Specifically we can enhance or suppress the myocardial relaxation by adjusting $P_\text{endo}$.  
Our results in Table \ref{tab::EDPEPLEffects}) show that, with an enhanced myocardial relaxation, say, when $P_\text{endo}\ge12$\,mmHg, there is more filling during diastole, compared to the cases when $P_\text{endo}<12$\,mmHg  under the same EDP.  This in turn gives rise to higher ejection fraction and stroke volume. However, if myocardial relaxation is suppressed, diastolic filling is less efficient, with subsequently smaller ejection fraction and stroke volume.  In the extreme case, when the myocardial relaxation is entirely absent, chamber volume increases by only 29.5 mL, and ejection fraction decreases to 29\%.  To maintain stroke volume obtained for $P_\text{endo}$=12\,mmHg, EDP needs to be as high as 20\,mmHg.  Indeed, increased EDP due to an impaired myocardial relaxation has been reported in a clinical study by Zile et al.~\cite{zile2004diastolic}.  A higher EDP indicates the elevated filling pressure throughout the refilling phase.  Increased filling pressure can help to maintain a normal filling volume and ejection fraction, but runs the risks of ventricular dysfunction in the longer term, because pump failure will occur if no other compensation mechanism exists. 

During diastole,  the MV-LV model seems to yield a smaller orifice  compared to the corresponding CMR images.  
In our previous study \cite{gao2014finite}, the MV was  mounted in a rigid straight tube, the peak diastolic filling pressure is around 10\,mmHg, and the peak flow rate across the MV is comparable to the measured value (600\,mL/s). While in this coupled MV-LV model, even though with additional $P_\text{endo}$, the peak flow rate (200\,mL/s) is much less than the measured value. One reason is because of the extra resistance from the LV wall, which is absent in the MV-tube model \cite{gao2014finite}. The diastolic phase can be divided into three phasse~\cite{nishimura1997evaluation}: the rapid filling, slow filling, and atrial contraction. During rapid filling, the transvalvular flow is resulted from myocardial relaxation (the sucking effect), which contributes to 80\% of the total transvalvular flow volume. During slow filling and atrial contraction, the left atrium needs to generate a higher pressure to provide additional filling. In the coupled MV-LV model, the ramped pressure in the top plane of the inflow tract during late-diastole is related to the atrial contraction, and during this time, only 10\% of the total transvalvular flow occurs. However, the peak flow rate in rapid filling phase is much lower compared to the measured value, which suggests the myocardial relaxation would be much stronger.

In a series of studies based on in vitro $\mu$CT experiments, Toma \cite{toma2015fluid,toma2016fluid,toma2016fluidFSI} suggested that MV models with simplified chordal structure would not compare well with experimental data, and that a subject-specific 3D chordal structure is necessary.  This may explain some of the discrepancies we observed here.  A  simplified chordal structure is used in this study because we are unable to reconstruct the chordal structure from the CMR data.   CT imaging may allow the chordae reconstruction but it comes with radiation risk. Patient-specific chordal structure in the coupled MV-LV model would require further improvements of in vivo imaging techniques.


Several other limitations in the model may also contribute to the discrepancies.  These include the uncertainty of patient-specific parameter identification, uncertainties in MV geometry reconstruction from CMR images, the passive response assumption around the annulus ring and the valvular region of the LV model, and the lack of pre-strain effects.  Studies addressing these issues are already under way.  
We expect that further improvement in personalized modelling and more efficient high performance computing would make the modelling more physiologically detailed yet fast enough for applications in risk stratification and optimization of therapies in heart diseases.

%


\section{Conclusion}
We have developed a first fully coupled MV-LV model that includes fluid-structure interaction as well as experimentally constrained descriptions of the soft tissue mechanics.  
The model geometry is derived from in vivo magnetic resonance images of a healthy volunteer. It incorporates three-dimensional finite element representations of the MV leaflets,  sub-valvular apparatus, and the LV geometry.  Fibre-reinforced hyperelastic constitutive laws are used to describe the passive response of the soft tissues, and the  myocardial active contraction is also modelled.    The developed MV-LV model is used to simulate MV dynamics, LV wall deformation, and ventricular flow throughout the cardiac cycle.  Despite several modelling limitations, most of the results agree with in vivo measurements.  We find that with impaired myocardial active relaxation, the diastolic filling pressure needs to increase significantly in order to maintain a normal cardiac output, consistent with clinical observations. The model thereby represents a further step towards a whole-heart multiphysics modelling with a target for clinical applications.

\section*{Acknowledgement}
We are grateful for the funding from the UK EPSRC (EP/N014642/1, and EP/I029990/1) and the British Heart Foundation (PG/14/64/31043), and the National Natural Science Foundation of China (No. 11471261). In addition,  Feng received the China Scholarship Council Studentship and the Fee Waiver Programme at the University of Glasgow, Luo is funded by a Leverhulme Trust Fellowship (RF-2015-510), and Griffith is supported by the National Science Foundation (NSF award ACI 1450327) and the National Institutes of Health (NIH award HL117063)

\section*{Conflict interests}
The authors have no conflicts of interest.

\section*{References}
\bibliography{mybibfile}

\newpage
\begin{table}
	\caption{Material parameter values for MV leaftlets, chordae and the myocardium}
	\label{tab::passiveMat}
	\begin{tabular}{ccccccccc}
		\hline
		\multicolumn{2}{c}{\textbf{MV leaflets}} &  \multicolumn{2}{c}{$C_1$ (kPa)}  &  \multicolumn{2}{c}{$a_\text{v}$ (kPa)}  &   
		\multicolumn{2}{c}{$b_\text{v}$ (kPa)} & \\ 
		\multicolumn{2}{c}{Anterior} & \multicolumn{2}{c}{17.4}  & \multicolumn{2}{c}{31.3} & \multicolumn{2}{c}{55.93}  & \\
		\multicolumn{2}{c}{Posterior} & \multicolumn{2}{c}{10.2}  & \multicolumn{2}{c}{50.0} & \multicolumn{2}{c}{63.48} & \\
		\hline
		\multicolumn{2}{c}{\textbf{Chordae}} & \multicolumn{2}{c}{$C$ (kPa)} & & & & & \\
		\multicolumn{2}{c}{systole} & \multicolumn{2}{c}{9000}  & & & &  & \\
		\multicolumn{2}{c}{diastole} & \multicolumn{2}{c}{540}  & & & &  & \\
		\hline
		{\textbf{Myocardium}} & \multicolumn{8}{c}{}  \\ 
		passive   & $a$ (kPa) & $b$ & $a_\text{f}$ (kPa) & $b_\text{f}$ & $a_\text{s}$ (kPa) & $b_\text{s}$ & $a_\text{8fs}$ (kPa) & $b_\text{8fs}$ \\
		& 0.24 & 5.08 & 1.46 & 4.15 & 0.87 & 1.6 & 0.3 & 1.3 \\
		active    & \multicolumn{2}{c}{$T^\text{req} = 225\,$kPa}& &  & & & & \\
		\hline
	\end{tabular}
\end{table}
\clearpage

\begin{table}[h]
	\footnotesize
	\caption{Effects of EDP and the endocardial pressure loading ($P_\text{endo}$) on MV and LV dynamics.}
	\label{tab::EDPEPLEffects}
	\begin{tabular}{ccccccc}
		\hline
		Cases (mmHg) & $t^\text{iso-con}$(ms) & $t^\text{ejection}$(ms) & $V_\text{LV}^\text{ejection}$(mL) & $V_\text{MV}^\text{filling}$(mL) & $F_\text{MV}^\text{peak}$(mL/s) & LVEF(\%) \\
		\hline
		EDP=8, $P_\text{endo}$=8  & 60 & 227 & 52   & 60.6 & 412.93 & 47\% \\
		EDP=8, $P_\text{endo}$=10 & 58 & 237 & 57.6 & 65.9 & 442.60 & 49\% \\ 
		\textbf{EDP=8, $P_\text{endo}$=12} & 57 & 243 & 63.2 & 72.1 & 468.41 & 51\% \\
		EDP=8, $P_\text{endo}$=14 & 55 & 251 & 67.8 & 76.8 & 486.84 & 53\% \\
		EDP=8, $P_\text{endo}$=16 & 54 & 256 & 72.3 & 81.3 & 503.76 & 54\% \\
		EDP=8, $P_\text{endo}$=0  & 75 & 174 & 20.8 & 29.5 & 209.54 & 29\% \\
		EDP=12,$P_\text{endo}$=0  & 64 & 213 & 41.0 & 50.9 & 343.47 & 42\% \\
		EDP=14,$P_\text{endo}$=0  & 61 & 226 & 50.6 & 59.8 & 406.81 & 47\% \\
		EDP=16,$P_\text{endo}$=0  & 58 & 243 & 61.8 & 71.9 & 459.64 & 51\% \\ 
		EDP=18,$P_\text{endo}$=0  & 56 & 251 & 68.5 & 79.3 & 486.58 & 54\% \\
		EDP=20,$P_\text{endo}$=0  & 55 & 262 & 75.7 & 86.2 & 511.16 & 55\% \\
		\hline 
	\end{tabular}
\end{table}

\newpage

\newpage

\newpage

\newpage

\newpage

\newpage


\end{document}